# Magnetization reversal in the orthovanadate $RVO_3$ compounds (R = La, Nd, Sm, Gd, Er and Y): Inhomogeneities caused by defects in the orbital sector of quasi-one-dimensional orbital systems

L D Tung*, M R Lees, G Balakrishnan, and D M$^c$K Paul

Department of Physics, University of Warwick, Coventry CV4 7AL, United Kingdom.

*Abstract:* We report on a study of various $RVO_3$ single-crystal samples (R = La, Nd, Sm, Gd, Er and Y) which show temperature-induced magnetization reversal. For compounds with lighter rare-earths (R = La, Nd and Sm), magnetization reversal can be observed for magnetic field applied in the *ab* plane and along the *c* axis, whereas for the heavy rare-earths (R = Gd, Er) magnetization reversal is only observed when the field is applied along the *a* axis. $YVO_3$ has a magnetization reversal along all the main crystallographic axes in a modest applied field. We have found that some compounds are very sensitive to small trapped fields present in the superconducting solenoid of the magnetometer during the cooling. Based on the observed results, we argue that inhomogeniety caused by defects in the orbital sector in the quasi one-dimensional orbital systems could account for the unusual magnetization reversal.

*PACS numbers: 75.60.Jk; 75.50.Ee; 42.50.Lc*

---

*Current address: Department of Physics, University of Liverpool, L69 7ZE, United Kingdom. Email: leductung2000@gmail.com

## I- INTRODUCTION

In a ferrimagnet, which contains two or more different types of magnetic ions, or one magnetic ion positioned at different crystallographic sites, the magnetization is partially cancelled due to the antiferromagnetic coupling of the (two) magnetic sublattices with unequal values of magnetization. In addition where there is a strong single-ion magnetic anisotropy, the magnetization direction is further fixed along a certain crystallographic axis. If the temperature dependence of the magnetization of the (two) antiferromagnetically coupled sublattices is different, a magnetization reversal (MR) can sometimes be observed. The temperature at which the total magnetization vanishes is called the compensation point $T_o$. This phenomenon of MR has been observed in many ferrites after it was first predicted by Néel [1].

Surprisingly MR has also been observed in some orthovanadate $RVO_3$ compounds (R = rare-earths), namely (polycrystalline) $LaVO_3$ [2,3], (single-crystal) $YVO_3$ [4,5], (polycrystalline) $SmVO_3$ [6], and (polycrystalline) $NdVO_3$ [7] when the field-cooled (FC) magnetization was measured in a modest applied magnetic field. In $LaVO_3$ and $YVO_3$, vanadium is the only magnetic constituent with almost equal V-V bonds, and thus all sites are magnetically equivalent. In $SmVO_3$ and $NdVO_3$, the rare-earth moments (i.e., Sm or Nd) remain disordered or order at temperatures much below the MR temperature of the compound. It would seem that the aforementioned mechanism for the ferrites does not appear to be applicable to the orthovanadate compounds.

In the case of $LaVO_3$, the most widely accepted explanation of the MR is due to Goodenough *et al.* [8,9] who have proposed the presence of canted spin antiferromagnetism as a result of the Dzyaloshinsky-Moriya (DM) antisymmetric exchange appearing below the antiferromagnetic spin ordering (SO) transition $T_{SO}$ =

143 K. They argued that on cooling, the response of the orbital moment to the forces generated by the first order magnetostrictive distortion, in which the crystallographic structure of the compound changes from orthorhombic to monoclinic symmetry at $T_o$ = 138 K, can reverse the Dzyaloshinsky vector so as to create a canted spin component in a direction opposite to the applied field below $T_o$. In $YVO_3$, a similar MR is observed at $T_o$ = 95 K when the material is FC with a modest field (≤ 1 kOe) applied along the *a* axis [4,5]. Since there is no structural change at $T_o$ in $YVO_3$, Ren *et al.* [4,5] proposed a mechanism in which a canted spin structure in the *ab* plane is driven by both the single–ion magnetic anisotropy and DM coupling. They assumed that these two spin canting mechanisms produce moments pointing in opposite directions and that the competition between these two mechanisms can lead to the observed MR at $T_o$. On further cooling, $YVO_3$ shows another, first order, MR at $T_s$ = 77 K. At $T_s$, the compound also experiences a first order change in the crystallographic structure from a monoclinic to an orthorhombic symmetry [10] accompanied by a change from a high temperature C-type (ferromagnetic along the *c* axis and antiferromagnetic in the *ab* plane) to a low temperature G-type (antiferromagnetic in all directions) magnetic structure [11,12].

The mechanism of opposed canted spins in which the spin quantum number S was used in the magnetic energy for $YVO_3$ [4,5] instead of the absolute value of thermal averaged spin vector $|\langle S \rangle|$ was, however, questioned by Kimishima *et al.* who argued that if $|\langle S \rangle|$ was used for the anisotropy energy, the canting angle of spin magnetic moment would become constant with a positive value, and that a negative magnetization should not be observed [7]. For (polycrystalline) $NdVO_3$, there is a single MR temperature at $T_o$ = 13.1 K which was then explained by Kimishima *et al.* based on the model of N-type ferrimagnetism [7]. The origin of the ferrimagnetism was attributed to an imbalance in the quenching rate of the orbital magnetic moments of the $V^{3+}$ ions [3]. The N-type ferrimagnetism model was also used to explain qualitatively the two MR events at $T_o$ = 122 K and $T_s$ = 68 K observed in (polycrystalline) $SmVO_3$ [6].

It is clear that to date, there is no consensus as to the mechanism behind the unusual MR observed in the various orthovanadate $RVO_3$ compounds. Therefore, it was the aim of this study to carry out a systematic investigation of different orthovanadate $RVO_3$ single-crystal samples.

## II- EXPERIMENTAL DETAILS

Single-crystals of $RVO_3$ (R = La, Nd, Sm, Gd, Er and Y) were grown by means of the floating zone technique using a high temperature xenon arc-furnace. At first, $RVO_4$ phase was prepared by mixing stoichiometric quantities of $R_2O_3$, and $V_2O_5$ (with purity of 99.9%) followed by annealing at temperature from 950-1100 $^oC$ for 48 hrs. The product was then reduced at 1000 $^oC$ in flowing $H_2$ for 10-24 hrs to create the polycrystalline $RVO_3$ phase. The $RVO_3$ feed and seed rods used for the single-crystal growth were then made by pressing the powder under hydrostatic pressure and annealing the rods at 1500 $^oC$ under a flow of Ar for 6 hrs. The crystals were checked using a back-scattering Laue technique which shows the sharp and consistent spots in different Laue pictures indicating the good quality. For the magnetic measurements, all crystals were being cut cube shape with dimensions ranging from about 1.5 to 2.5 mm along the main crystallographic *a*, *b* and *c* axes according to the *Pbnm* orthorhombic lattice. We would like to note that, for $LaVO_3$ and $NdVO_3$ since the lattice parameters *a* and *b* are very close to each other (*a* = 5.5178 Å, *b* = 5.5149 Å for $LaVO_3$ and *a* = 5.451 Å, *b* = 5.575 Å for $NdVO_3$), it is difficult to recognize these axes from the conventional Laue technique as well as to rule out the possibility of twining in the *ab* plane of the crystals. On the other hand, the lattice parameter *c* is much larger than *a* and *b* (*c* = 7.7897 Å for $LaVO_3$ and *c* = 7.74 Å for $NdVO_3$) and thus the *c* direction can be identified unambiguously and that the crystals of these two compounds can be seen untwined along this direction. For other $RVO_3$ (R = Sm, Gd, Er and Y) compounds, the crystals are untwined and all the axes can be identified unambiguously from the Laue technique.

The heat capacity measurements were performed in the temperature range 1.8-300 K using a Quantum Design Physical Property Measurement System (PPMS). The *zero*-field-cooled (*ZFC*) (Ref. [13]) and FC measurements of the magnetization were carried out in a Quantum Design SQUID magnetometer. In the FC measurements, the sample is cooled from the paramagnetic region to 1.8 K in an applied field with the data taken either on cooling (FCC) or on warming (FCW) with the same magnetic field applied as on cooling. For the *ZFC* measurements, the sample was cooled in *zero* field to 1.8 K before the magnetic field is applied. The data are then taken on warming. For the field dependence of the magnetization, we used either the SQUID magnetometer or an Oxford Instruments vibrating sample magnetometer (VSM).

**III- RESULTS**

In Fig. 1, we present the heat capacity versus temperature data for the different $RVO_3$ compounds. For $LaVO_3$, we observe a sharp peak at a temperature of 136 K then a shoulder at about 139 K. This behaviour is similar to that reported by other groups who observe a peak in the heat capacity at 141 K and a shoulder at 143 K [14,15]. We therefore identify the peak at 136 K as the orbital ordering (OO) temperature $T_{OO}$ and the shoulder at 139 K as the antiferromagnetic SO temperature $T_{SO}$. The fact that our transition temperatures are about 4-5 K lower than those reported earlier could be attributed to a difference in the stoichiometric compositions of the samples.

For $YVO_3$, the heat capacity results shown in Fig. 1 point to an OO transition at $T_{OO}$ = 199 K which is then followed by an antiferromagnetic SO transition at $T_{SO}$ = 115 K. We note that when cooling the sample in the SO region, there is a sudden drop in the heat capacity at a temperature $T_s$ = 77 K. Previously, Blake *et al.* [10] and Miyasaka *et at.* [15] have also reported a drop in the heat capacity of $YVO_3$ at this temperature, although the drop in their data is much smaller than in ours. We believe that the observed drop may not be (entirely) intrinsic to $YVO_3$ but is largely due to the fact that the sample becomes partially physically decoupled from the *sample stage* during the measurement. Strain induced micro-cracks in the sample, which are sometimes visible with an optical microscope, are the most likely source of this decoupling. Although we are not able to resolve this problem, we show a few data points just below $T_s$ in order to indicate the position of the transition. The transition at $T_s$ is also seen in our magnetic measurements. We note that the values of the three transition temperatures observed in our $YVO_3$ crystal are consistent with those reported by Ren *et al.* [4,5]. For $ErVO_3$, we observe a behaviour in the heat capacity results that is similar to that seen in $YVO_3$ (see Fig. 1). The OO transition occurs at $T_{OO}$ = 190 K, followed by antiferromagnetic SO at $T_{SO}$ = 110 K and $T_s$ = 49 K. For $GdVO_3$ and $SmVO_3$, the values of $T_{OO}$ and $T_{SO}$ as shown in Fig. 1 are 199 K, 118 K and 192.6 K, 130 K, respectively. In these two compounds, at low temperature, we observe an additional peak at a temperature defined as $T_M$ of about 8 K ($GdVO_3$) and 12 K ($SmVO_3$). This transition at $T_M$ is evident in our magnetic measurements (see below) for $GdVO_3$ but it is not visible in the magnetization data for $SmVO_3$. For $NdVO_3$, an OO occurs at $T_{OO}$ = 183 K, with an antiferromagnetic SO at $T_{SO}$ = 136 K. We see no further transition in the heat capacity data in the SO region down to 1.8 K. For this compound, recent neutron diffraction results have suggested that there is no spontaneous magnetic order on the neodymium sublattice, but that the neodymium moments are gradually polarized by the ordered vanadium moments below 60 K [12].

In Fig. 2, we present the FCC and FCW magnetization versus temperature data for $LaVO_3$ measured along different crystallographic axes at different applied fields. Earlier, we already mentioned that $LaVO_3$ has lattice parameter *a* and *b* almost equal to each other, and it is not possible to recognize the *a* and *b* axes from the conventional Laue technique. As a result, we would refer the notations *a* and *b* in this Figure for $LaVO_3$ compound as the two main axes in the *ab* plane which, as a result of twinning, might have an alternation of *a* and *b* axes. MR can be observed when a modest field is applied along the *c* axis and in the *ab* plane. The MR temperature, $T_o$, appears to be field dependent with a value of about 133 K and 129.5 K for applied fields of 0.1 and 1 kOe, respectively. Another characteristic feature of the data is the splitting of the FCC and FCW curves, which become very obvious in, for instance, an applied field of 4 kOe (see Fig. 2, right panels). We note that this feature is generally considered to be a fingerprint of a first order phase transition. For $LaVO_3$, it was known of the first order phase transition at $T_{OO}$ [14]. However, the feature of splitting between FCC and FCW is very unusual in the sense that it spreads over quite a large temperature range and becomes particularly obvious at certain applied fields (e.g., 4 kOe).

The FCC and FCW magnetization versus temperature curves for $NdVO_3$ are presented in Fig. 3. Again similarly to the case of $LaVO_3$, we note that in this Figure, the notations *a* and *b* should be understood as the two main axes in the *ab* plane which, as a result of twinning, might have an alternation of *a* and *b* axes. At low applied fields (10 and 100 Oe), MR can be observed along the *c* axis and in the *ab* plane at a temperature $T_o$ of about 15 K which is slightly higher than the value of 13.1

K reported for a polycrystalline sample [7]. With increasing applied field, MR can still be observed at an applied field of 1 kOe in the *ab* plane. Along the *c* axis, the FCC curve shows two MR temperatures. In contrast, for the FCW curve, the magnetization becomes positive over the entire temperature range studied. When the applied field is 4 kOe, the MR disappears, although there is a dip in the curves, which also occurs at 15 K. We note that neutron diffraction results for the $NdVO_3$ compound have shown no changes in the crystallographic structure at $T_o$ [12]. In addition, there has not been known any first order phase transition in the SO region for the compound. However, like it is the case for $LaVO_3$, in $NdVO_3$ we also observe a splitting of the FCC and FCW over a wide temperature range that becomes particularly obvious at, for instance, an applied field of 1 kOe (see Fig. 3, right panels).

In contrast to $LaVO_3$ and $NdVO_3$, $SmVO_3$ has two MR temperatures $T_o$ and $T_s$ where the magnetization smoothly changes in sign (Fig. 4). At low applied fields, e.g., of 10 or 100 Oe, $T_s$ = 63.8 K and $T_o$ = 127.5 K is observed compared to $T_s$ = 68 K and $T_o$ = 122 K reported for a polycrystalline sample [6]. At 1 kOe, MR can still be observed for a field applied along the *a* axis, but not for fields applied along the *b* and *c* axes. With a large enough applied field (50 kOe), the magnetization eventually becomes positive over the entire temperature range. We also note that there are no features visible in the magnetization data that correspond with the peak observed at 12 K in the heat capacity of $SmVO_3$. It, therefore, seems reasonable to identify this peak in the heat capacity as a Schottky anomaly arising from a combination of crystal field effects and the samarium ions.

In Fig. 5, we present the FCC and FCW data for $GdVO_3$ measured in different applied fields. MR can be observed in modest applied fields (e.g., 10 and 100 Oe) only along the *a* axis at temperatures $T_s \approx 25.5$ K and $T_o \approx 78$ K. In addition, we also observe a peak at $T_M \approx 8$ K along all three principal axes. This peak is also seen in the heat capacity data in Fig. 1 and is thus most probably related to an ordering of the gadolinium moments. When the magnetic field is large enough, e.g., 1 kOe, MR is no longer observed (see Fig. 5, right panels).

The FCC and FCW magnetization data for $ErVO_3$ are presented in Fig. 6. With a magnetic field of 100 Oe applied along the *a* axis, one can observe MR at two temperatures $T_s$ and $T_o$ in the FCW curve only. The MR transition at $T_s$ is first order, while at $T_o$, it is second order. While there is no MR along the *b* and *c* axes, a first order anomaly at $T_s$ can clearly be observed.

For $YVO_3$, earlier investigations have indicated that there are two MR temperatures in a modest field of ≤ 1 kOe directed along the *a* axis. These two MR temperatures are (almost) field independent. MR at $T_s = 77$ K is first order while MR at $T_o = 95$ K is second order [4,5,16]. The results of our measurements on a $YVO_3$ single-crystal sample at 1 kOe (Fig. 7, right panels) are consistent with previous studies in that MR can be observed only with a field along the *a* axis. However, measurements at lower field, e.g., 100 Oe or 10 Oe (Fig. 7, left panels), reveal that MR can be observed along the *b* and *c* axes as well.

One common feature of the orthovanadate $RVO_3$ compounds revealed by this investigation is that the results of a nominal *ZFC* magnetization versus temperature run appear to depend on the residual or trapped field (TF) in the magnetometer solenoid. Previously, we have already highlighted this special feature in $GdVO_3$ [17], and $PrVO_3$ [18]. A small TF in a superconducting magnet is inevitable unless the magnet is quenched. We have examined the effects of the TF carefully. Before each measurement, we ran a degauss sequence to minimize the TF, producing a residual field with an absolute value that is estimated to be less than 2 Oe. We can "*generate*" a TF with the opposite sign by reversing the signs of the magnetic fields in the degauss sequence (Ref. [19]). In Fig. 8, we present the results for *ZFC* magnetization versus temperature runs for various $RVO_3$ compounds where the data was collected in a positive field of 100 Oe after having been cooled in a positive trapped field (PTF) and a negative trapped field (NTF). For "*conventional*" magnetic materials domain translation is generally reversible at (very) low magnetic field. As a result, the TF does not usually have any influence on the *ZFC* data. This is clearly not the case for these orthovanadate compounds. What is surprising is that even though the absolute magnitude of the TF is small and is about two orders of magnitude less than the applied measuring field, the change in sign of the TF has led to an almost mirror behaviour in the *ZFC*_PTF and *ZFC*_NTF for compounds such as $NdVO_3$, $SmVO_3$ and $YVO_3$.

To investigate further the magnetic properties, we have measured the magnetization versus field along the main axes of these $RVO_3$ compounds. The results for $LaVO_3$, $NdVO_3$, $SmVO_3$ and $YVO_3$ are displayed in Fig. 9 and for $GdVO_3$ and $ErVO_3$ in Fig. 10. For $LaVO_3$, $NdVO_3$, $SmVO_3$ and $YVO_3$, there is the hysteresis at 1.8 K along the *c* axis and in the *ab* plane. This hysteresis remains up to temperatures near $T_{SO}$ (data not shown). On the other hand, for $GdVO_3$ and $ErVO_3$, there is virtually no hysteresis in the magnetization curves measured at 1.8 K (Fig. 10, left panel). However, hysteresis is evident at higher temperature, e.g., 50 K for $GdVO_3$ and 65 K for $ErVO_3$ (Fig. 10, right panels), particular for fields applied along the *a* axis.

## IV- DISCUSSIONS

From the results of the present investigation, it can be seen that there is a clear evolution in the behaviour of the $RVO_3$ compounds as the radius of the rare-earth element changes. For compounds containing rare-earths with a large ionic radius (R = La, Nd and Sm), MR can be observed for fields applied along the *c* axis and in the *ab* plane, whereas MR can only be observed for fields along the *a* axis in compounds with rare-earths that have a smaller radius (R = Gd and Er). It is interesting to note that Y has a similar radius to Gd, thus the $YVO_3$ compound lays close to the borderline between these two regimes.

Similarly, one can see that compounds with R = La, Nd, Sm and Y have open hysteresis loops along the *c* axis and in the *ab* plane (Fig. 9). On the other hand, for $GdVO_3$, such features are only observed along the *a* axis at elevated temperatures (Fig. 10), which is correlated with MR only along this direction. For $ErVO_3$ at 65 K, an open hysteresis loop is most obvious for a field applied along the *a* axis compared with the *b* axis where the remanent magnetization is very small, and with the *c* axis where there is virtually no hysteresis. This behaviour may also be linked with the fact that MR is observed only along the *a* axis for this compound. It appears that the appearance of a MR is correlated with the (unusual) development of coercivity and remanent magnetization in these antiferromagnetic materials. This latter feature has to date been attributed to a homogeneous canted antiferromagnetism, which can produce a net moment (see e.g., [4,5,8,9]).

For $LaVO_3$, canted antiferromagnetism has not been seen in neutron diffraction data. Early powder neutron diffraction studies by Zubkov *et al.* [20] have shown that the compound has a simple collinear C-type antiferromagnetic structure. Our recent single-crystal neutron diffraction data [21] confirms the results of Zubkov *et al.* In addition, based on our magnetization data, it is clear that the MR temperature $T_o$ is not only field dependent, but also it occurs well below $T_{OO}$ where the first order structural change from an orthorhombic to a monoclinic symmetry begins to take place [14]. Thus, based on the experimental observations, we feel that a mechanism for MR which requires a canted antiferromagnetism and magnetostriction, as proposed by Goodenough *et al.* [8,9] may not be applicable to the case of $LaVO_3$.

In the case of $YVO_3$, Ren *et al.* [4,5] proposed a small canting angle of ~0.2$^o$ of magnetic moments with the spins lying in the *ab* plane for a C-type antiferromagnetic ordering above $T_s$ and in the *ac* plane for a G-type antiferromagnetic ordering below $T_s$. Their proposal, however, does not appear to be confirmed by the recent single-crystal neutron data from Ulrich *et al.* [11] and Reehuis *et al.* [12] who showed that below $T_s$ the compound has simple collinear G-type antiferromagnetic structure and above $T_s$ the moments are canted by an angle $\theta = 16.5^o$ out of the *ab* plane. This means that experimentally the canting of the spins at $T_s < T < T_{SO}$ is very different from that expected from the model based on opposed canted spins. In addition, the collinear G-type antiferromagnetic structure at $T < T_s$ would indicate that the observed remanent magnetization and coercivity of the compound at 1.8 K (Fig. 9d) does not originate from a weak ferromagnetism arising from the canting of the spins. Our further investigation along the *b* and *c* axes indicates that MR can also be observed along these directions in a (very) small field, e.g., 10 and 100 Oe. Thus, the spin canting mechanism, which can give only an *a* axis ferromagnetic component [4,5,22] does not appear to be able to account for MR in the orthovanadate $RVO_3$ compounds including R = La, Nd, Sm and Y where MR can be observed along the *c* axis as well.

Further evidence against the use of a spin canting mechanism to explain for MR seen in these materials is based on the sensitivity to low magnetic fields as characterized in Fig. 8. We have found that this low field sensitivity is common to many of the other orthorvanadates $RVO_3$ (R = Pr, Ce, Lu) compounds that do not exhibit MR (see e.g., in Ref. [18]). In other published work, for instance on $CeVO_3$ [23,24], the *ZFC* magnetization curve was reported to start with a negative value at the lowest temperature and then become positive as the temperature is increased. This observation was attributed to a MR in the compound. However, if the material is cooled in absolutely zero field, the "perfect" ZFC curve should always start with a positive value for the magnetization at the lowest temperature (i.e., the magnetization is aligned coincident with the magnetic field applied after the cooling). Our detailed investigation for the orthovanadate $RVO_3$ compounds as shown in Fig. 8 clearly indicate cooling in a residual field of only a few Oe can significantly influence the results of the nominal *ZFC* magnetization. Whilst it would be interesting to explore in detail the origin of the low field sensitivity of these orthovanadate $RVO_3$ compounds, we believe that this special feature can also be taken as evidence against the scenario of homogeneous opposed canted spins. Within the framework of this model, a fit of the magnetization data leads to a very large single-ion anisotropy as well as DM interactions of the order of 25 meV [5]. This value is about two orders of magnitude higher than the 0.33 meV derived experimentally from recent neutron inelastic scattering results [11]. It also seems unreasonable that the competition between the antiferromagnetic interactions, the single-ion anisotropy, and the DM interactions, characterized by the very large energy of 25 meV can be influenced by such small trapped fields.

As we mentioned earlier, another mechanism proposed to explain the MR, due to Kimishima *et al.*, is based on a model of N-type ferrimagnetism [6,7]. This mechanism is, however, not supported by the results of neutron diffraction measurement on various $RVO_3$ compounds, which have shown no difference between the vanadium moments at different sites [11,12,20,21,24-26].

What is the mechanism for MR in the orthovanadate $RVO_3$ compounds? To address this question we would like to reconsider the open hysteresis loops, which to date have been assumed to result from a homogeneous canted antiferromagnetism (see e.g., [4,5,8,9]). Taking a different approach, we note that such open hysteresis loops can also arise as a result of inhomogeneities producing some random field (RF) spins that coexist with spins in an infinite antiferromagnetic network. These RF spins will give rise to a finite coercivity and a remanent magnetization [27]. Such a scenario has been observed in some diluted antiferromagnet Ising systems [28,29]. Actually, some of the diluted antiferromagnet Ising systems, e.g., $(CH_3NH_3)Mn_{1-x}Cd_xCl_3.2H_2O$ [30], exhibit magnetic properties, including MR and a sensitivity to low magnetic fields, that are very similar to those of the orthovanadate $RVO_3$ compounds investigated here. In the othovanadate $RVO_3$ compounds, the one-dimensional (Ising) orbital character has been well established [11,31-33]. We argue that defects in the orbital sector (faulty orbitals) in the orthovanadate $RVO_3$ compounds can act in a way that is similar to the substitution of a non-magnetic for a magnetic element in the diluted antiferromagnet Ising systems and such defects lie at the origin of the RF spins. To the best of our knowledge, there have been no neutron diffuse scattering experiments to detect the presence of the RF spins for any of the orthovanadate $RVO_3$ compounds. However, other experimental probes such as optical conductivity studies on $LaVO_3$ [34] or phonon activity studies on $YVO_3$ [35] have already suggested that inhomogeneity may play an important role in these materials. In the case of $YVO_3$, Massa *et al.* [35] suggested that the presence of faulty orbitals in the material is connected with lattice inhomogeneity. A model of inhomogeneity with RF spins was also supported in the case of $PrVO_3$ where one can observe the staircase-like hysteresis loop and spin glass-like behaviour of the *ZFC*FC curves [18] which are similar to those seen in some diluted antiferromagnet Ising systems [36,37]. For $PrVO_3$, it was argued that faulty orbitals and thus the presence of the RF spins may also arise as a result of orbital quantum fluctuations which might be a characteristic feature of the orthovanadate $RVO_3$ compounds because of their relatively weak Jahn-Teller interactions [11,12]. In anyway, regardless of the source of the inhomogeneity, the similar MR and low field sensitivity of the orthovanadate $RVO_3$ compounds and some of the diluted Ising antiferromagnets like $(CH_3NH_3)Mn_{1-x}Cd_xCl_3.2H_2O$ [30] do suggest the same physical mechanism based on the inhomogeneity with RF spins scenario. In this sense, it also becomes more plausible to us that there is indeed a link between the special feature of MR and that of quasi one-dimensional orbital character in the orthovanadate $RVO_3$ compounds.

Based on the model of inhomogeneity with RF spins, the compounds can be viewed as consisting of weak local field spins (sublattice A) embedded in the main matrix of strongly antiferromagnetically coupled spins (sublattice B). We note that this picture does not rule out the possibility that a (long-range) canted antiferromagnetic structure is present in sublattice B. In other words, the canted antiferromagnetic structures found by neutron diffraction for some compounds such as $YVO_3$ (only at intermediate temperature region $T_s < T < T_{SO}$) [11,12], $CeVO_3$ [24], $LuVO_3$ [25], $YbVO_3$ [26] are not in conflict with this scenario. In addition, the inhomogeneity with RF model is different, but not inconsistent with, the coexistence, at low temperature, of a monoclinic and an orthorhombic phase found recently in a high resolution x-ray powder diffraction study of $SmVO_3$ [38]. To avoid complications arising due to the contribution from the rare-earth magnetism, we estimate the percentage of the RF spins only for the compounds with non magnetic rare-earths (i.e., $LaVO_3$ and $YVO_3$). At low temperature, the value of the saturation magnetization, $M_s$, is 1.3 $\mu_B$/f.u. [20] and 1.72 $\mu_B$/f.u. [11,12] for $LaVO_3$ and $YVO_3$, respectively. Given that the antiferromagnetic structures of both $LaVO_3$ and $YVO_3$ are collinear at 1.8 K, the observed remanent magnetization should be only due to the RF spins. A rough estimate based on the ratio $M_r/2M_s$ suggests only 0.21% and 0.23% of the vanadium moments in $LaVO_3$ and $YVO_3$, respectively contribute to this signal. The very small percentage of these disorder spins could explain why they have not been detected by some microscopic techniques such as the neutron experiments [11,12,20,21,24-26]. We note that, in chain-like Ising antiferromagnets, the inclusion of impurities in concentrations of less than 1 % can seriously affect the overall magnetic properties [30,39]. This occurs because the introduction of impurities limits the growth of the correlation length along the chains, dramatically influencing the long-range magnetic ordering and the magnetic response of the spins within each chain [30]. In addition, it is also worth noting that the majority of spins in the $LaVO_3$ and $YVO_3$ systems (sublattice B) are indeed strongly antiferromagnetically coupled. For instance, based on the high field susceptibility of $LaVO_3$ and $YVO_3$ we estimate that the field needed to fully polarize the magnetic moments would be of the order of $5 \times 10^3$ kOe. The Weiss temperatures ($\theta_p$) of the orthovanadate compounds also indicate a strong antiferromagnetic coupling, with both $LaVO_3$ and $YVO_3$ having the Weiss temperature $\theta_p < -100$ K for fields applied along all three principal crystallographic axes (data not shown). This means that for most measuring fields, the magnetic susceptibility of the main antiferromagnetic phase (sublattice B) is very small and thus it can be dominated by the minority phase (sublattice A) which has spins with much weaker local field and that causes MR as observed for different compounds.

## V- CONCLUSIONS

In summary, we have studied MR in several orthovanadate $RVO_3$ compounds (R = La, Nd, Sm, Gd, Er and Y). We have found that there is a clear evolution in the behaviour of this series of compounds that is correlated with changes in the rare-earth radius. Compounds with light rare-earths (R = La, Nd and Sm) exhibit a MR along the *c* axis and in the *ab* plane, whereas for compounds with heavy rare-earths (R = Gd and Er) MR is observed only for magnetic fields applied along the *a* axis. Some compounds are very sensitive to a small cooling field, which affects the behaviour of the nominal *ZFC* data. Based on the results of our systematic studies, we suggest that the physics behind MR in the orthovanadate $RVO_3$ compounds may be linked to inhomogenieties caused by a (very) small amount of defects in the orbital sector in the quasi one-dimensional orbital systems.

## ACKNOWLEGEMENTS

One of the authors, LDT, would like to thank Prof. A. Paduan-Filho for his kindness in sending useful information regarding the issue of MR in the diluted antiferromagnetic materials and Prof. R.W. McCallum, Dr. J.P. Goff for invaluable discussions. The support from EPSRC grant (UK) is gratefully acknowledged.

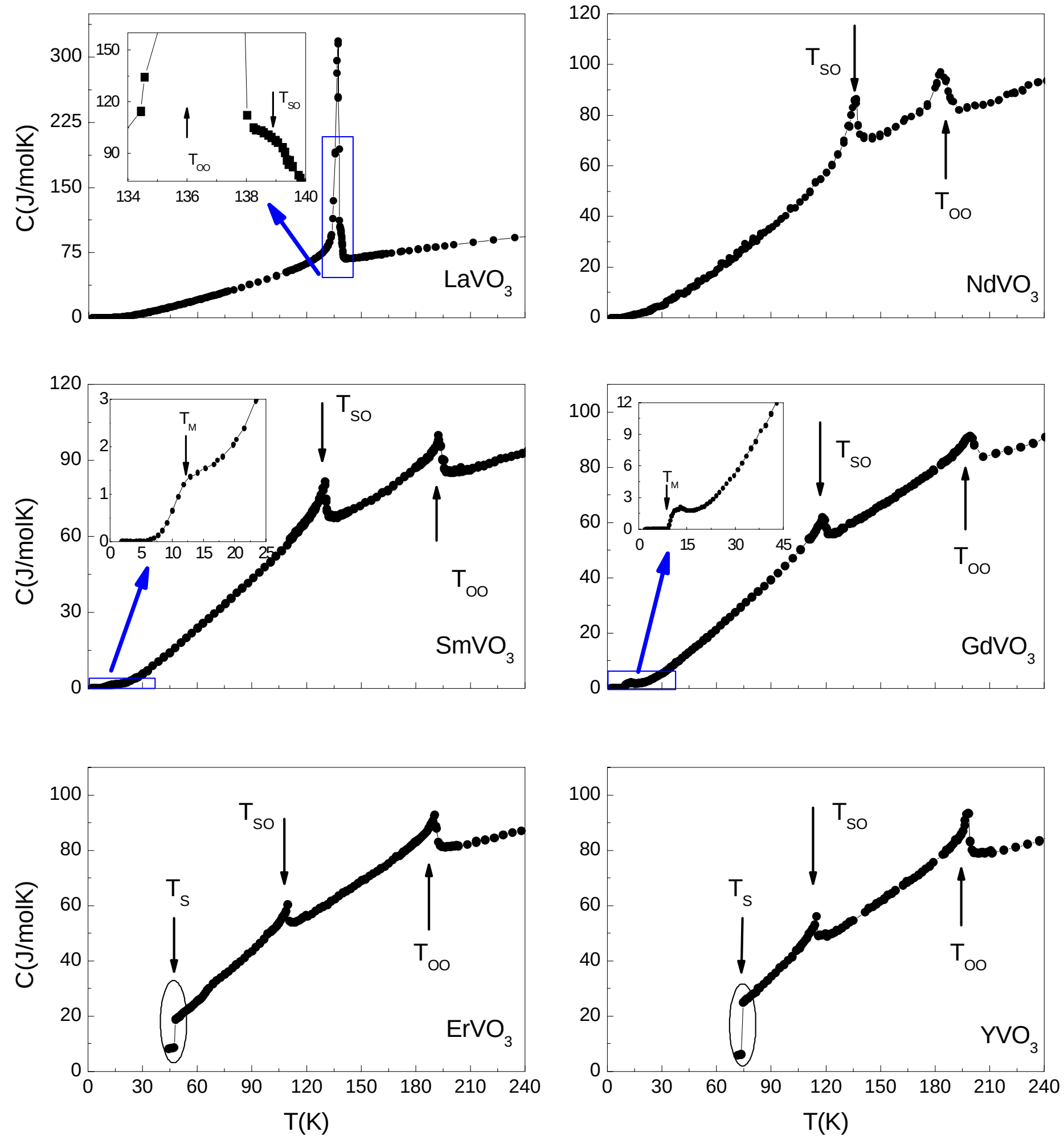

Fig. 1: Heat capacity as a function of temperature of various orthovanadate $RVO_3$ crystals as indicated. The definition of transition temperatures $T_{OO}$, $T_{SO}$, $T_s$, and $T_M$ can be found in the text.

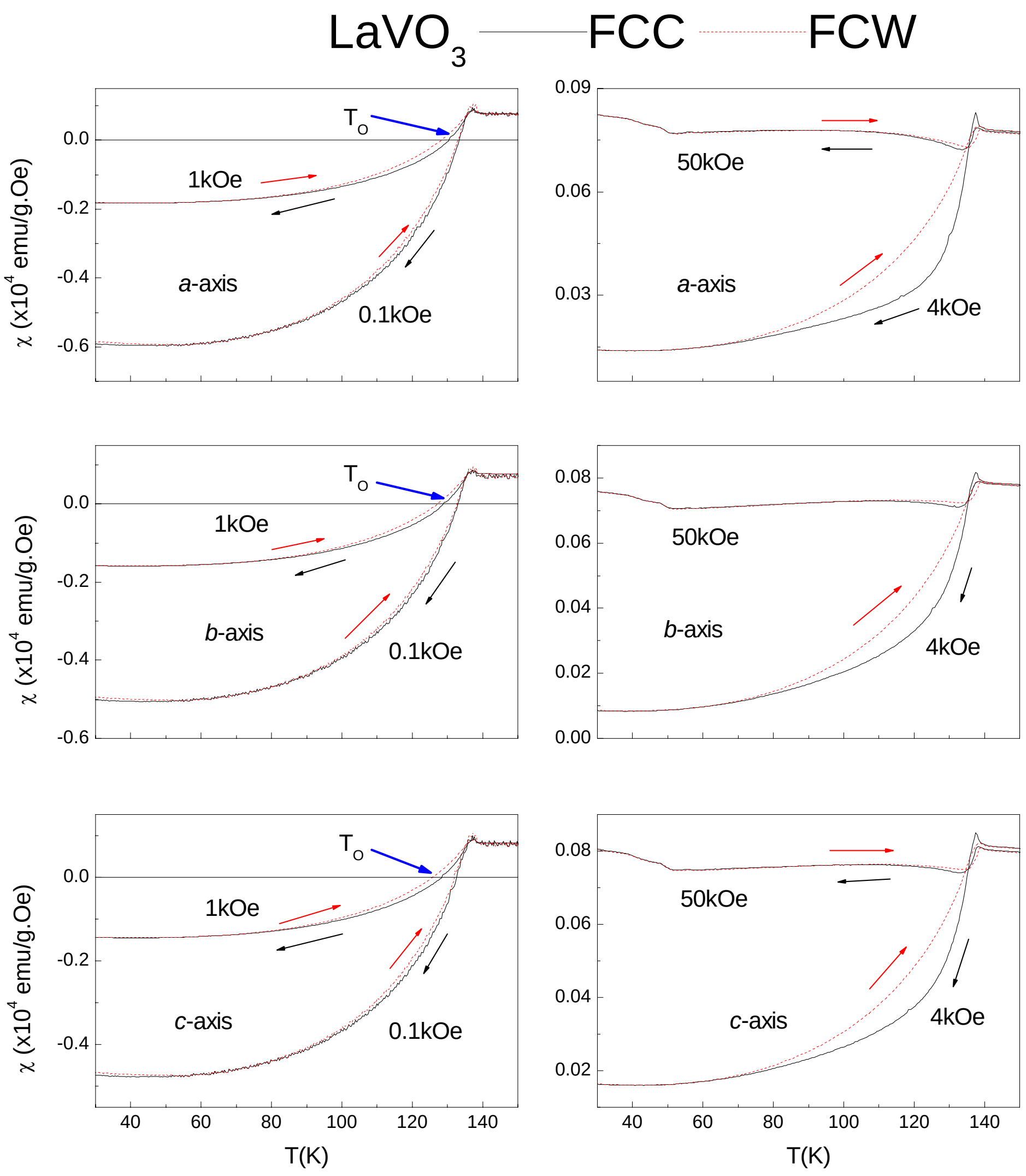

Fig. 2 (color online): FCC (solid lines) and FCW (dashed lines) measured along the main axes of a $LaVO_3$ crystal in different applied fields as indicated.

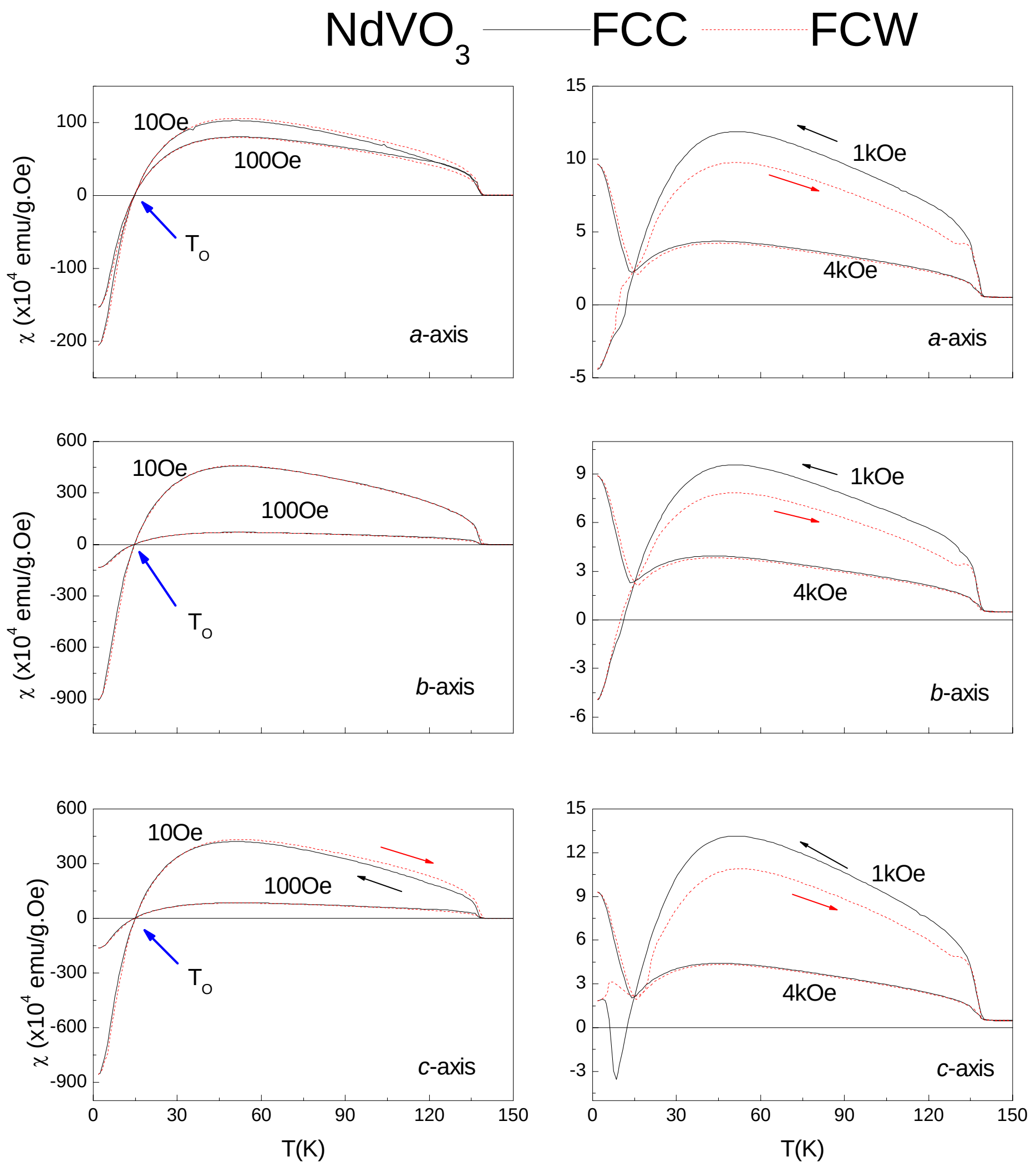

Fig. 3 (color online): FCC (solid lines) and FCW (dashed lines) measured along the main axes of a $NdVO_3$ crystal in different applied fields as indicated.

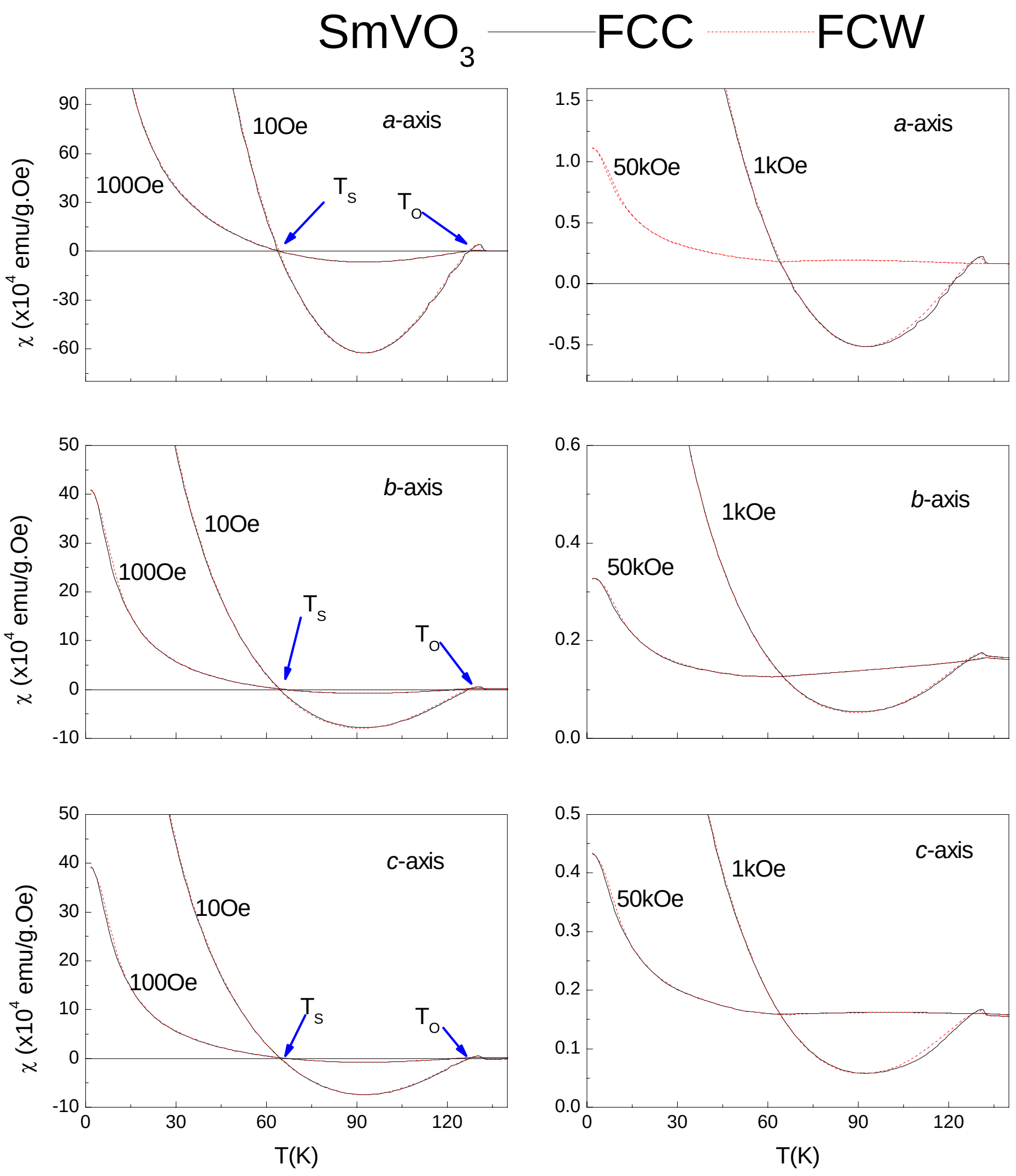

Fig. 4 (color online): FCC (solid lines) and FCW (dashed lines) measured along the main axes of a $SmVO_3$ crystal in different applied fields as indicated.

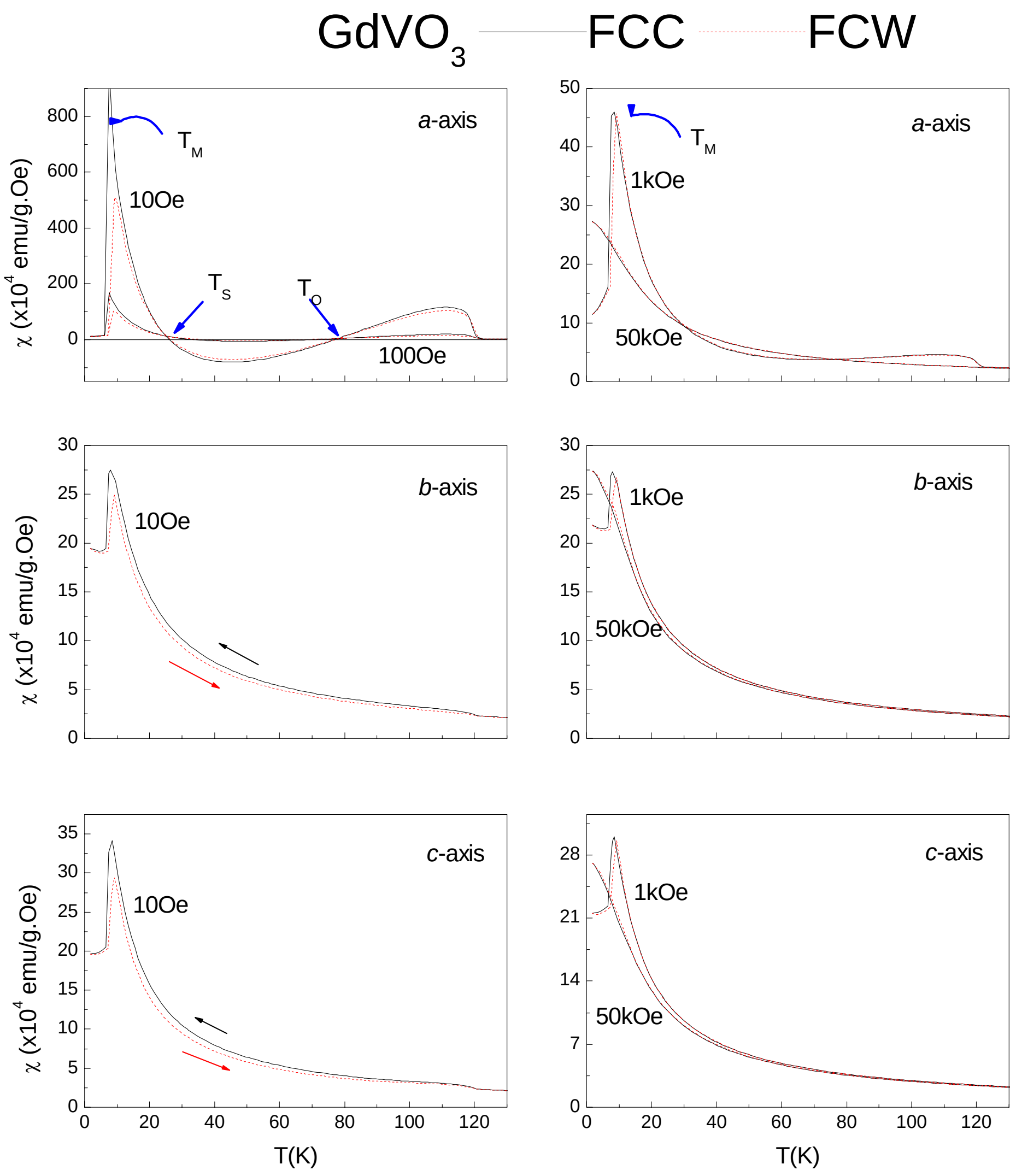

Fig. 5 (color online): FCC (solid lines) and FCW (dashed lines) measured along the main axes of a $GdVO_3$ crystal in different applied fields as indicated.

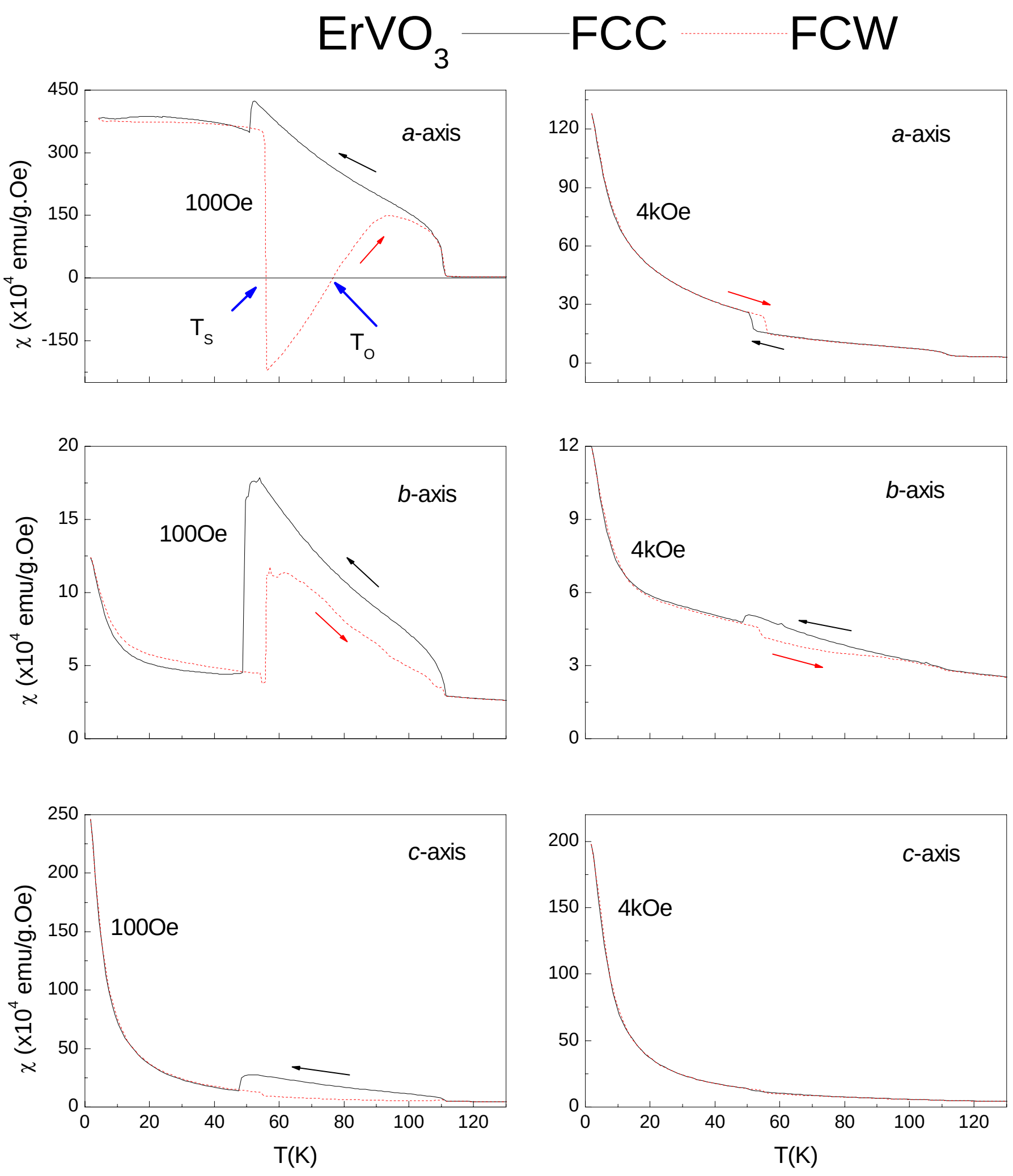

Fig. 6 (color online): FCC (solid lines) and FCW (dashed lines) measured along the main axes of a $ErVO_3$ crystal in different applied fields as indicated.

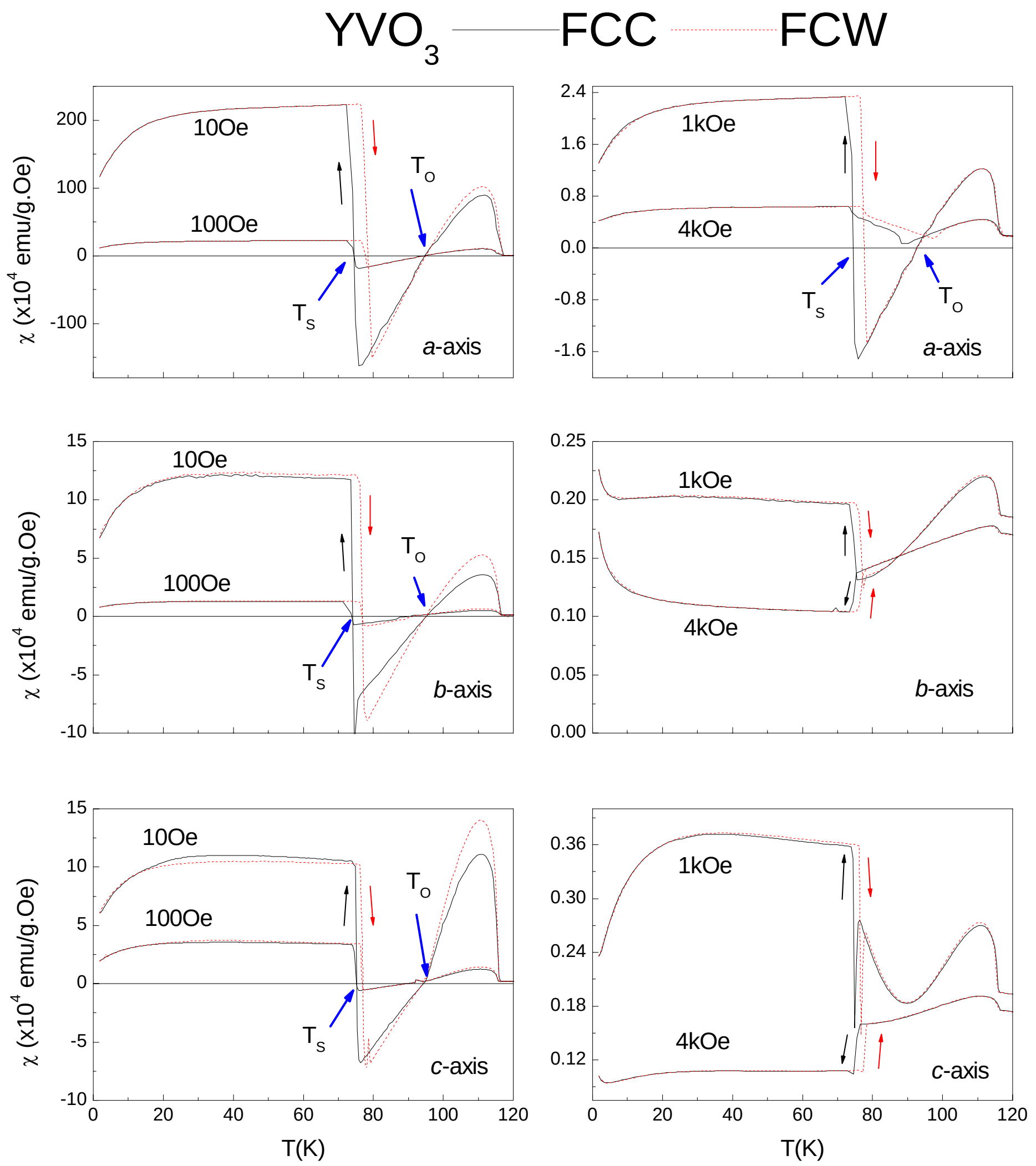

Fig. 7 (color online): FCC (solid lines) and FCW (dashed lines) measured along the main axes of a $YVO_3$ crystal in different applied fields as indicated.

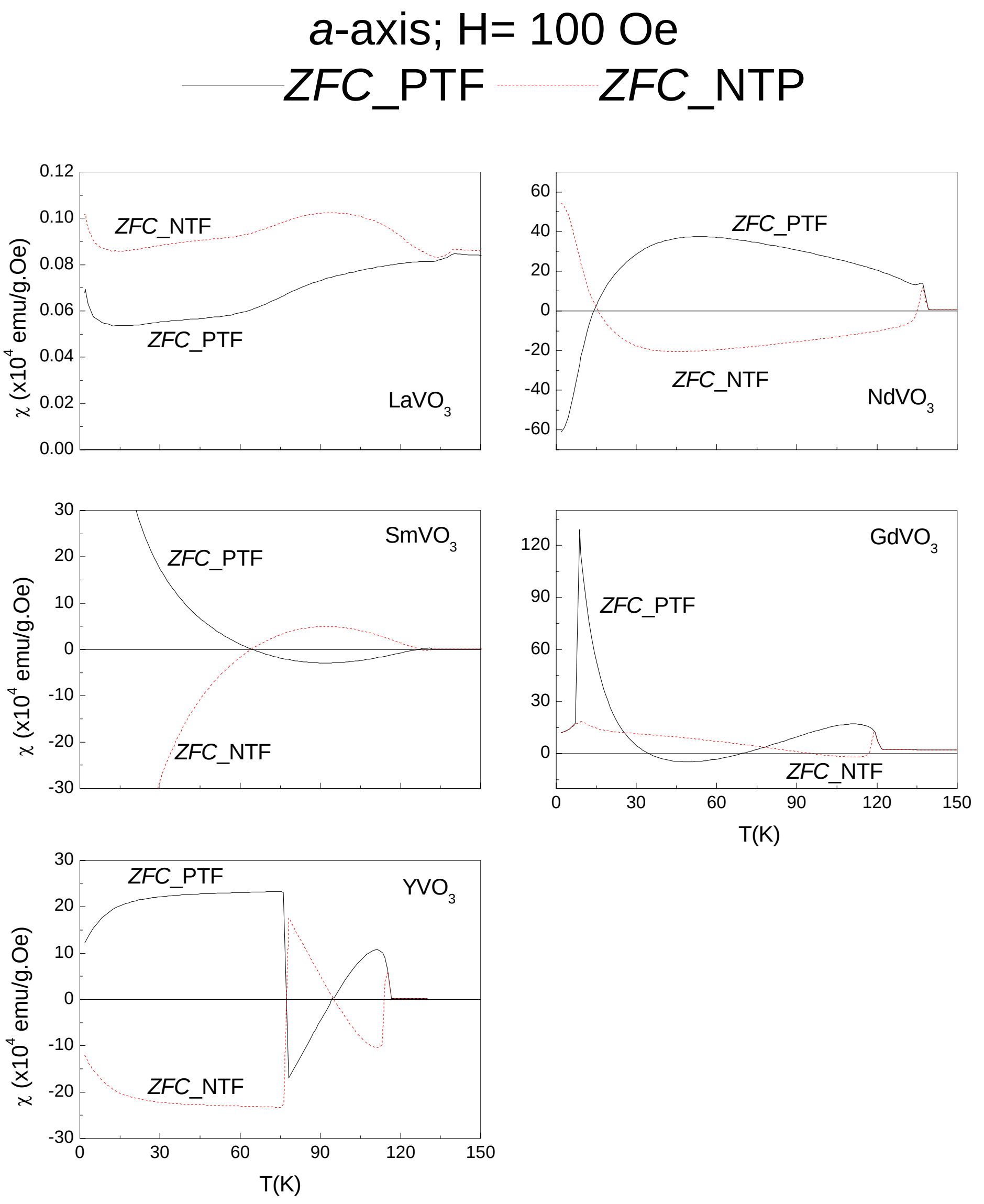

Fig. 8 (color online): Difference between results of the *ZFC* bound with positive trapped field (*ZFC*_PTF (solid lines)) and with negative trapped field (*ZFC*_NTF (dashed lines)) measured along the *a* axis for various orthovanadate $RVO_3$ compounds as indicated.

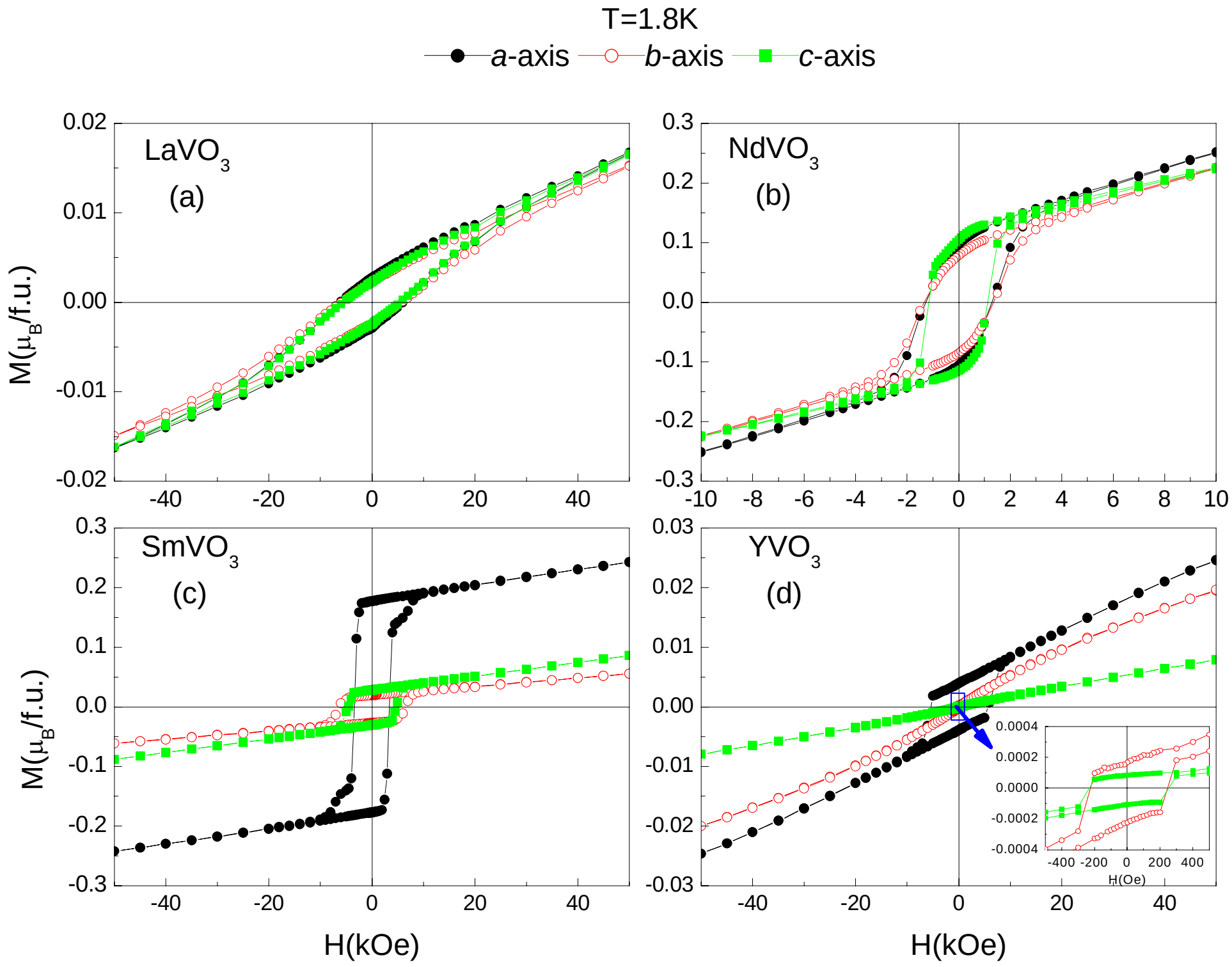

Fig. 9 (color online): The magnetization as a function of magnetic field (hysteresis) measured along the main axes of $LaVO_3$, $NdVO_3$, $SmVO_3$ and $YVO_3$ at T = 1.8 K. The inset in Fig. 9d shows an extended view of the hysteresis loop around the origin for $YVO_3$.

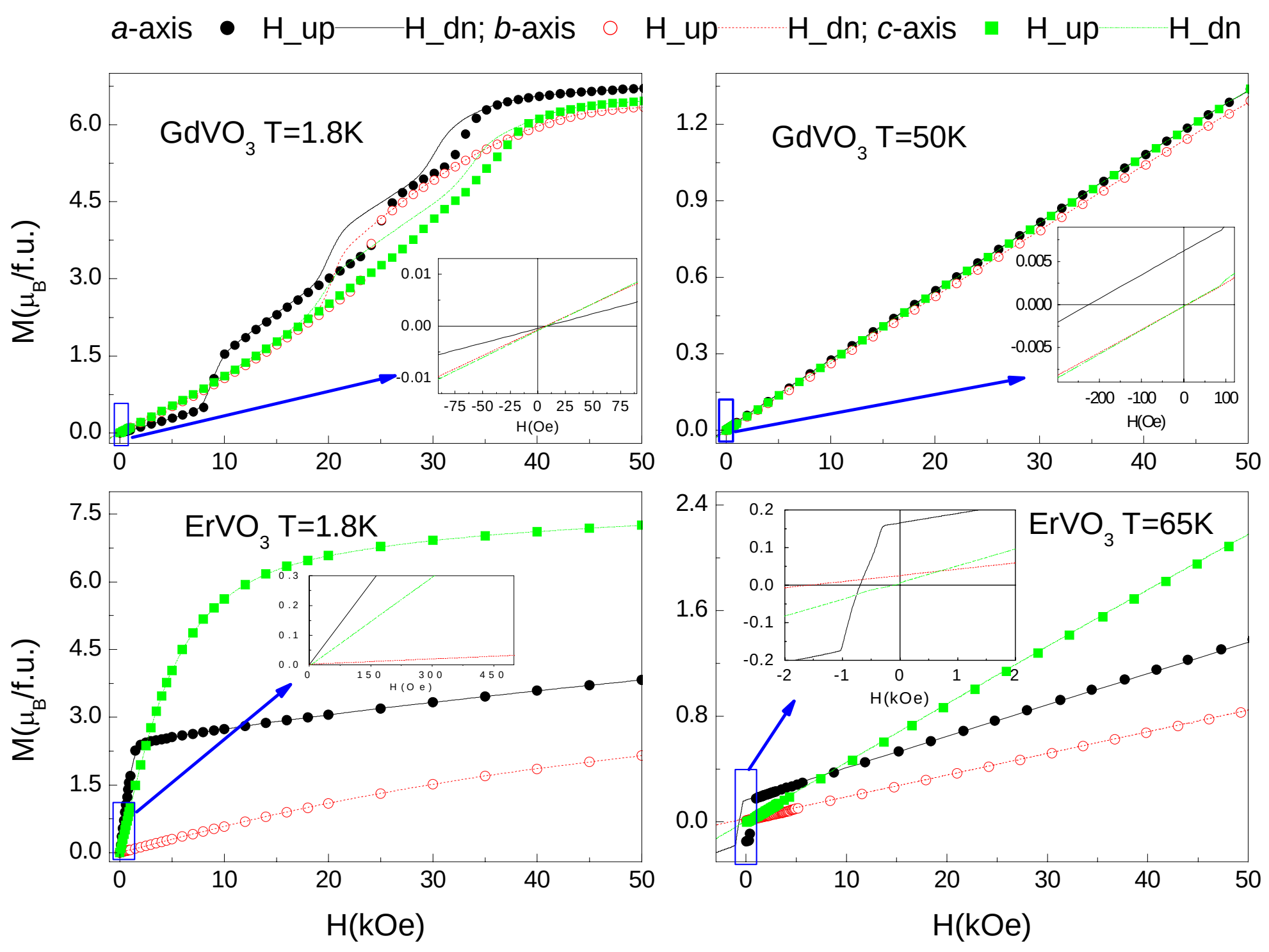

Fig. 10 (color online): The magnetization as a function of magnetic field (hysteresis) measured along the main axes at different temperatures as indicated for $GdVO_3$ (top panels) and $ErVO_3$ (bottom panels). The inset shows an extended view of the decreasing section of the magnetic isotherm around the origin.